\documentclass[twocolumn,raggedbottom]{aastex701}

\usepackage{amsmath}

\begin{document}

\title{Roughness-Dependent Radiative Effects on Dimorphos within Hera’s Reach}


\author[orcid=0009-0004-6339-282X,gname=Yixuan,sname=Wu]{Yixuan Wu}
\altaffiliation{These authors contributed equally to this work as co-first authors.}
\affiliation{Tsinghua University, Beijing 100084, China}
\email[]{}

\author[orcid=0009-0004-6155-4602,gname={Kin Thong},sname=Lee]{Kin Thong Lee}
\altaffiliation{These authors contributed equally to this work as co-first authors.}
\affiliation{Tsinghua University, Beijing 100084, China}
\email[]{}

\author[orcid=0000-0003-1097-0521,gname=Yifei,sname=Jiao]{Yifei Jiao}
\affiliation{Tsinghua University, Beijing 100084, China}
\affiliation{University of California, Santa Cruz, CA 95064, USA}
\email[]{}

\author[orcid=0000-0002-8025-9113,gname=Bin,sname=Cheng]{Bin Cheng}
\affiliation{Tsinghua University, Beijing 100084, China}
\email[show]{bincheng@tsinghua.edu.cn}

\author[gname=Zhaokui,sname=Wang]{Zhaokui Wang}
\affiliation{Tsinghua University, Beijing 100084, China}
\email[]{}

\author[gname=Hexi,sname=Baoyin]{Hexi Baoyin}
\affiliation{Tsinghua University, Beijing 100084, China}
\affiliation{Inner Mongolia University of Technology, Hohhot 010051, China}
\email[]{}

\begin{abstract}
Following the Double Asteroid Redirection Test (DART) impact, radiative recoil, tides, and other dissipative processes drive the dynamical evolution of the Didymos–Dimorphos binary, yet the effect of unresolved surface roughness on radiative migration remains poorly quantified.
We incorporate this roughness into a binary thermophysical model coupled to planar spin--orbit dynamics.
During the six-month investigation by ESA's Hera mission, surface roughness significantly reduces the radiative contribution to the mean orbital contraction rate by $28$--$39\%$ relative to smooth surfaces for two post-impact shape models of Dimorphos in synchronous principal-axis rotation.
With surface roughness included, planar-libration calculations still yield centimeter-scale radiation-driven orbital contraction that increases mean motion, contributing a mean of $40^\circ$--$85^\circ$ to orbital-phase advance from DART to Hera's arrival.
During Hera's investigation, radiative recoil in these calculations contributes a mean orbital-phase advance of $0.7^\circ$--$2.0^\circ$, which could be detectable given Hera's projected $0.05^\circ$ formal orbital-longitude uncertainty.
In the modeled tumbling state, the net radiative contribution to the mean orbital contraction rate over the same interval is much smaller, with surface roughness reducing it by $16$--$18\%$.
These results identify thermal roughness as a key factor in radiative migration estimates that, combined with Hera’s orbital and rotational observations, can constrain post-impact dissipation and rubble-pile mechanical properties relevant to planetary defense.
Refining the thermophysical treatment of roughness with Hera’s thermal and shape measurements can improve binary YORP estimates from remote infrared observations of other synchronous binaries, helping infer tidal response from measured orbital migration to probe asteroid interiors and advance understanding of binary evolution and lifetimes.
\end{abstract}

\keywords{\uat{Asteroid dynamics}{2210} --- \uat{Asteroid satellites}{2207} --- \uat{Orbital evolution}{1178} --- \uat{Asteroid surfaces}{2209} --- \uat{Asteroid rotation}{2211}}


\section{Introduction}\label{sec:introduction}

The lifetimes and diverse configurations of small binary asteroids are shaped by the evolution of their mutual orbits under radiative forces and tides, together with episodic perturbations from planetary encounters and impacts \citep{walsh2015formation,dai2026multigeneration}.
In particular, solar radiation transfers momentum through absorption, reflection, and thermal re-emission, producing forces that can drive orbital migration at rates comparable to or greater than those produced by tides \citep{vokrouhlicky2015yarkovsky,jacobson2011equilibria}.
For an asymmetric, synchronously rotating secondary, the resulting radiative recoil can exert a secular torque on the mutual orbit through the binary Yarkovsky--O'Keefe--Radzievskii--Paddack (BYORP) effect.
BYORP can expand the orbit toward escape or shrink it until the components collide on $10^4$--$10^5$ yr timescales, much shorter than near-Earth asteroid dynamical lifetimes \citep{cuk2005effects,mcmahon2010secular}.
A balance between inward BYORP and outward primary tides can prolong survival and help explain the abundance of binaries \citep{jacobson2011equilibria,cueva2026survival}.

However, the radiative contribution to these evolutionary pathways remains uncertain.
For 1999~KW$_4$, the BYORP expansion rate predicted from the radar-derived secondary shape exceeds the measured net rate, raising questions about the adopted shape and the treatment of radiative recoil \citep{scheirich2021drift}.
Beyond the shape-dependent BYORP torque, eclipses and thermal irradiation from the companion can drive migration through the binary Yarkovsky effect, as thermal inertia delays the surface temperature response to these changes in heating.
This mechanism can act on an asynchronously rotating primary (BYP), even with a synchronous secondary contributing BYORP, and can also act on an asynchronously rotating secondary (BYS) \citep{zhou2024binary,zhou2024yarkovsky}.

Surface roughness below the resolution of a shape model further modifies surface temperatures through local shadows and self-heating, while unequal temperatures on inclined surfaces redirect the escaping radiation \citep{rozitis2011directional,davidsson2015interpretation}.
The resulting thermal beaming can change predicted heliocentric Yarkovsky drift and YORP spin acceleration by tens of percent \citep{rozitis2012influence}.
For binary systems, BYORP is sensitive to relief explicitly represented in secondary shape models \citep{cuk2024byorp}, but the additional effect of unresolved thermal roughness remains poorly quantified \citep{sorli20253d}.

The Didymos--Dimorphos system offers an opportunity to investigate how thermal roughness modifies the radiative contribution to orbital evolution.
For the pre-impact system, thermophysical modeling of ground-based infrared observations constrained the primary's thermal roughness \citep{rozitis2024preimpact}, and long-term orbital monitoring showed that the mutual orbit was contracting slowly, consistent with a near balance between BYORP and tides \citep{scheirich2022preimpact,cueva2024secular}.
NASA's Double Asteroid Redirection Test (DART) impact shortened Dimorphos's orbital period by approximately 33 minutes \citep{thomas2023orbital} and may have globally reshaped the secondary without leaving a distinct crater \citep{raducan2024physical}.
Subsequent orbit determinations suggest that the period shortened by a further $34\pm15$~s over the following weeks \citep{naidu2024orbital}.
This additional shortening is unlikely to arise from interactions with bound ejecta \citep{agrusa2025gravitational} and may instead reflect eccentricity damping through internal dissipation in Dimorphos \citep{cuk2024byorp}.
Mutual-event timings in 2025 also hint at an orbital phase advance relative to earlier predictions \citep{knight2025ground}.

Radiative-recoil estimates that account for Dimorphos's rotational state help constrain post-impact dissipation when combined with orbital and rotational observations from ESA's Hera mission.
\citet{zhou2024yarkovsky} predicted BYS-driven migration detectable by Hera for a modeled asynchronous post-impact state.
Dimorphos's post-impact spin state remains unresolved: modeling of the early rotational lightcurves favors non-principal-axis tumbling that is synchronous on average \citep{pravec2024rotational}, while a principal-axis libration model successfully reproduces the orbital observations \citep{naidu2024orbital,richardson2024dynamical}.
Even slow orbital migration driven by radiative recoil and tides changes the orbital period, allowing appreciable phase shifts to accumulate \citep{cueva2024secular}.
Hera's forthcoming measurements of both bodies' masses, shapes, and spatially resolved thermal emission will inform these recoil estimates, complementing its observations of the system's orbital and rotational evolution \citep{michel2022esa,okada2025tiri,meyer2023energy}.

Here we quantify how unresolved thermal roughness changes the radiative contribution to orbital migration for representative post-impact shapes and rotational states of Dimorphos.
The planar solution of \citet{naidu2024orbital} provides a detailed set of orbital and physical parameters constrained by observations, which we adopt together with the infrared-derived thermophysical constraints of \citet{rozitis2024preimpact}.

\section{Methods}  \label{sec:methods}

Thermophysical calculations for the synchronously rotating secondary of 1996~FG$_3$ have quantified how thermal inertia, eclipses, and mutual heating modify BYORP \citep{sorli20253d}.
Our calculations incorporate unresolved thermal roughness into the validated thermophysical–dynamical framework of \citet{wu2026iac}, implemented in COBIRD \citep{wu2026cobird}.
Heat conduction and radiative exchange determine the radiative forces and torques, which feed back on the mutual orbit and Dimorphos's rotation in the coupled planar integrations.

\begin{figure*}[t]
\centering
\includegraphics[trim=6.12bp 0 16.74bp 0,clip,width=0.95\textwidth]{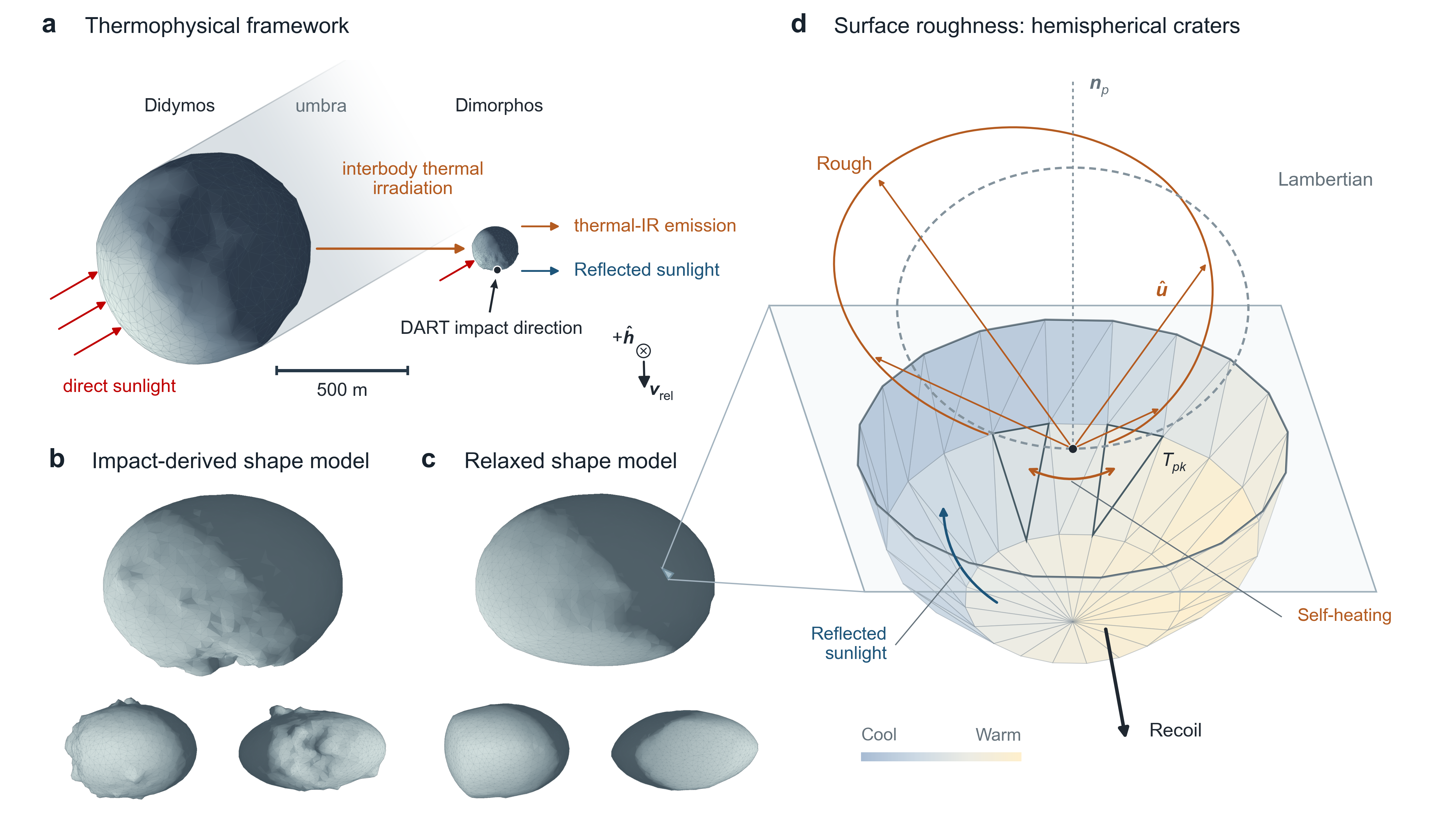}
\caption{Binary geometry, resolved shapes, and unresolved thermal roughness.
(a) Geometry from the post-impact orbital solution (JPL s547), radiation paths, and reconstructed DART approach, with $\hat{\boldsymbol{h}}$ denoting the orbit normal and $\boldsymbol{v}_{\mathrm{rel}}$ the relative velocity.
(b, c) Impact-derived and relaxed shape models viewed perpendicular to the orbital plane (top), with Didymos-facing and leading hemispheres at bottom left and right, respectively.
(d) Schematic roughness crater and emission profiles, with $T_{pk}$ denoting wall temperature, $\boldsymbol{n}_p$ the parent-facet normal, and $\hat{\boldsymbol{u}}$ an emission direction.
Radial distance represents thermal power per solid angle, with equal total hemispheric power for the orange profile and dashed gray Lambertian reference.
Crater-wall colors indicate relative temperature, and the black arrow shows net radiative recoil.
Red, orange, and blue denote direct sunlight, thermal radiation, and reflected sunlight, respectively.}
\label{fig:model-overview}
\end{figure*}

\subsection{Binary Thermophysics}
\label{subsec:binary-thermophysical}

Each shape facet carries a body-fixed, one-dimensional thermal column \citep{rozitis2011directional}, whose temperature profile evolves continuously with the changing illumination and attitude \citep{wu2026iac}.
With depth $z$ increasing inward and surface temperature \mbox{$T_s=T(0,t)$}, each column satisfies
\begin{equation}
 \begin{aligned}
 \rho C_p\frac{\partial T}{\partial t}
 &=\kappa\frac{\partial^2 T}{\partial z^2},\\[3pt]
 -\kappa\left.\frac{\partial T}{\partial z}\right|_0
 &=Q^{\mathrm{abs}}-\epsilon\sigma T_s^4,\\[3pt]
 \left.\frac{\partial T}{\partial z}\right|_{z_b}&=0.
 \end{aligned}
 \label{eq:thermal-column}
\end{equation}
Here $\rho$, $C_p$, $\kappa$, and $\epsilon$ denote density, specific heat capacity, thermal conductivity, and infrared emissivity, while $\sigma$ is the Stefan--Boltzmann constant and $z_b$ is the column depth.
Both bodies use the thermophysical parameters in Table~\ref{tab:model-parameters}, including the constant meteorite-based specific heat capacity \citep{rozitis2024preimpact,opeil2010thermal}.
The adopted thermal inertia for particulate regolith varies with heliocentric distance $r_h$ as $\Gamma\propto r_h^{-3/4}$ \citep{rozitis2024preimpact}, with normalization at $1.1\,{\mathrm{au}}$ and thermal conductivity $\kappa=\Gamma^2/(\rho C_p)$.
The thermal model uses the nominal Dimorphos bulk density \citep{daly2023successful,raducan2024physical}, whereas dynamical masses are taken from the independent s547 orbital solution.

The absorbed flux combines direct sunlight, sunlight reflected once from other facets (single-scattered sunlight), and thermal irradiation,
\begin{equation}
 \begin{aligned}
 Q_i^{\mathrm{abs}}={}&(1-A_{\mathrm{B}})(Q_i^\odot+Q_i^{\mathrm{sc}})\\
 &+\epsilon(Q_i^{\mathrm{IR}}+Q_i^{\mathrm{bin}}).
 \end{aligned}
 \label{eq:absorbed-flux}
\end{equation}
Here $A_{\mathrm{B}}$ is the Bond albedo, and $Q_i^{\mathrm{IR}}$ and $Q_i^{\mathrm{bin}}$ are the thermal irradiation from the same body and from its companion, respectively.
Direct sunlight is $Q_i^\odot=S_\odot\mu_i/(r_h/{\mathrm{au}})^2$, where $S_\odot$ is solar irradiance at 1 au and $\mu_i$ is the nonnegative incidence cosine reduced by shadows and eclipses.
Within each body, view factors $V_{ij}$ evaluated with the M2 approximation of \citet{rezac2020accuracy} convert outgoing flux at $j$ to incident flux at $i$, accounting for geometry and visibility.
For smooth facets, $Q_j^{\mathrm{ref}}=A_{\mathrm{B}}Q_j^\odot$ and
\begin{equation}
 \begin{aligned}
 Q_i^{\mathrm{IR}}&=\sum_j V_{ij}\epsilon\sigma T_{j,s}^4,\\[3pt]
 Q_i^{\mathrm{sc}}&=\sum_j V_{ij}Q_j^{\mathrm{ref}}.
 \end{aligned}
 \label{eq:terrain-heating}
\end{equation}
Thermal irradiation from Didymos contributes to Dimorphos's recoil through anisotropic re-emission of the absorbed energy \citep{sorli20253d,zhou2024yarkovsky}, with the incident flux on a smooth receiving facet evaluated using the center-line approximation of \citet{wu2026iac},
\begin{equation}
 Q_i^{\mathrm{bin}}=\frac{\chi_i\cos\alpha_i}{\pi r^2}\,L_{\mathrm{proj}},
 \label{eq:binary-heating}
\end{equation}
where $\chi_i$ describes companion visibility, $\alpha_i$ is incidence angle, $r$ is center separation, and $L_{\mathrm{proj}}$ is the primary's thermal flux integrated over its visible projected area as seen from Dimorphos.
Emission angles, incidence angles, and visibility are evaluated along the line of centers using the common separation $r$.
The much weaker reverse heating is neglected \citep[Section~5.1]{jackson2024thermophysical}, as are interbody sunlight scattering and multiple visible-light scattering.

For Lambertian thermal emission and diffuse reflection, the outgoing power per unit area is $q_i=\epsilon\sigma T_{i,s}^4+Q_i^{\mathrm{ref}}$ and the recoil is
\begin{equation}
 \boldsymbol{F}_i=-\frac{A_iq_i}{\pi c}
 \int_{\Omega_{i,\mathrm{esc}}}
 (\boldsymbol{n}_i\!\boldsymbol{\cdot}\!\hat{\boldsymbol{u}})\hat{\boldsymbol{u}}\,\mathrm{d}\Omega .
\label{eq:macro-recoil}
\end{equation}
Here $A_i$, $\boldsymbol{n}_i$, and $\hat{\boldsymbol{u}}$ are facet area, outward unit normal, and emission direction, and $c$ is the speed of light.
Emission and reabsorption within the same body transfer no net momentum to it, so $\Omega_{i,\mathrm{esc}}$ includes only directions above the emitting body's terrain horizon \citep{rozitis2012influence,yan2019yorp}, giving $\boldsymbol{F}_i=-2A_iq_i\boldsymbol{n}_i/(3c)$ for an unobstructed facet.
The sums of $\boldsymbol{F}_i$ and $(\boldsymbol{r}_i-\boldsymbol{r}_X)\boldsymbol{\times}\boldsymbol{F}_i$, with facet centers $\boldsymbol{r}_i$ and mass center $\boldsymbol{r}_X$, give the force and torque on body $X$ and are rotated into the dynamical frame each thermal step.
Companion radiation contributes to heating, but its incident photon momentum is omitted from the force calculation.
Specular reflection and direct solar-photon momentum transfer are omitted following \citet{sorli20253d}, the latter having no secular contribution in classical orbit-averaged BYORP theory \citep{steinberg2011binary}.

\subsection{Thermal Roughness}
\label{subsec:thermal-roughness}

Unresolved thermal roughness is represented by hemispherical craters covering a fraction $f_R=(\theta_{\mathrm{rms}}/49^\circ)^2$ of each shape facet (macrofacet $p$) \citep{rozitis2011directional,rozitis2012influence}, where $\theta_{\mathrm{rms}}$ is the model rms slope \citep{rozitis2024preimpact}.
Rough cases use the same coverage on both bodies, adopting the Didymos-dominated infrared estimate as a reference for Dimorphos, whereas smooth cases set $f_R=0$ but retain resolved relief and terrain self-heating.

The model retains the hemispherical-crater parameterization and 100-subfacet discretization of \citet{rozitis2024preimpact} for consistency with their thermal-infrared roughness estimate, without refining the global shape mesh.
Each crater wall has its own solar incidence, shadowing, and thermal column (Equation~\eqref{eq:thermal-column}), coupled to other walls by thermal radiation and single-scattered sunlight, while terrain shadows and eclipses are evaluated at the parent facet.
Under the one-dimensional heat-conduction approximation, geometrically similar craters produce the same recoil per unit opening area, so no absolute crater diameter is prescribed \citep{rozitis2012influence}.

The directional thermal source combines visible projected areas with the wall surface temperatures $T_{pk}$,
\begin{equation}
 \begin{aligned}
 w_k(\hat{\boldsymbol{u}})
 &=\frac{a_k^{\mathrm{vis}}(\hat{\boldsymbol{u}})
 (\boldsymbol{n}_k\!\boldsymbol{\cdot}\!\hat{\boldsymbol{u}})_+}{A_o\mu_u},\\[3pt]
 E_p^{\mathrm{dir}}(\hat{\boldsymbol{u}})
 &=\sum_k w_k(\hat{\boldsymbol{u}})\epsilon\sigma T_{pk}^4.
 \end{aligned}
 \label{eq:rough-directional-source}
\end{equation}
Here $k$ indexes crater walls with outward unit normals $\boldsymbol{n}_k$, $A_o$ is the area enclosed by the polygonal rim of the crater mesh, and $\mu_u=\hat{\boldsymbol{u}}\boldsymbol{\cdot}\hat{\boldsymbol{z}}>0$, with $\hat{\boldsymbol{z}}$ along the parent normal and $(x)_+=\max(0,x)$.
With directions and normals expressed in the crater frame, the visible wall area $a_k^{\mathrm{vis}}$ determines the fraction $w_k$ of the projected opening occupied by wall $k$, and $E_p^{\mathrm{dir}}/\pi$ gives the corresponding radiance.
The resulting thermal beaming reflects both geometry and wall-temperature contrasts \citep{rozitis2011directional,rozitis2012influence}.

Thermal exchange between macrofacets on the same body uses these weights for both emission and reception,
\begin{equation}
 \begin{aligned}
 \bar E_{p\to q}={}&(1-f_R)\epsilon\sigma T_{p,s}^4\\
 &+f_R E_p^{\mathrm{dir}}(\hat{\boldsymbol{u}}_{pq}),\\[3pt]
 Q_{qk}^{\mathrm{ext}}={}&\frac{1}{\eta_k}\sum_p V_{qp}
 w_k(\hat{\boldsymbol{u}}_{qp})\bar E_{p\to q}.
 \end{aligned}
 \label{eq:rough-directional-exchange}
\end{equation}
Here $\eta_k=a_k/A_o$ is the ratio of the full area $a_k$ of wall $k$ to the crater-opening area $A_o$, and $\hat{\boldsymbol{u}}_{pq}$ points from emitting macrofacet $p$ to receiver $q$, with directions expressed in the appropriate crater frame.
The source $\bar E_{p\to q}$ combines smooth and rough emission, while $V_{qp}$ accounts for terrain visibility \citep{wu2026iac}.
Dividing by $\eta_k$ converts the flux assigned to wall $k$ per unit opening area into the incident flux per unit wall area, $Q_{qk}^{\mathrm{ext}}$, which enters the surface balance with absorptivity $\epsilon$, while heating within each crater is added separately using the same M2 formulation.
Single-scattered sunlight uses the same geometry with a reflected source and absorptivity $1-A_{\mathrm{B}}$, whereas interbody heating uses directions along the line of centers.
Unlike the sequential shape-facet and rough-surface solutions of \citet{rozitis2013global}, the exchange is updated from the evolving directional crater-wall emission.

With $Q_{pk}^{\mathrm{ref}}=A_{\mathrm{B}}Q_{pk}^{\odot}$ and $q_{pk}=\epsilon\sigma T_{pk}^4+Q_{pk}^{\mathrm{ref}}$, the rough recoil contribution for a parent facet of area $A_p$ is
\begin{equation}
 \begin{split}
 \boldsymbol{F}_p^{\mathrm{rough}}={}&-\frac{A_p}{\pi c}\mathbf R_p\sum_k q_{pk}\\
 &\times\int_{\mu_u>0} w_k(\hat{\boldsymbol{u}})\mu_u\,
 \mathcal V_p(\hat{\boldsymbol{u}})\hat{\boldsymbol{u}}\,\mathrm{d}\Omega,\\
 \boldsymbol{F}_p={}&(1-f_R)\boldsymbol{F}_p^{\mathrm{smooth}}
 +f_R\boldsymbol{F}_p^{\mathrm{rough}}.
 \end{split}
 \label{eq:rough-recoil}
\end{equation}
The rotation $\mathbf R_p=[\boldsymbol{x}_p,\boldsymbol{y}_p,\boldsymbol{n}_p]$ maps the crater frame to the body frame, with $\boldsymbol{x}_p$ and $\boldsymbol{y}_p$ orthonormal tangents to the parent facet.
The visibility $\mathcal V_p$ excludes photons intercepted by resolved terrain, while $w_k$ accounts for obstruction within the crater.
The parent-facet centroid defines the torque lever arm, while the directional weights and recoil integral use $1^\circ$ azimuth--elevation spacing, with solid-angle weighting for the integral.

Fixed body-frame geometry lets us precompute terrain-exchange, crater-self-heating, and recoil coefficients, avoiding repeated visibility calculations and angular recoil integrals.
Interbody weights follow the changing orbit and attitudes, while each wall's temperature and directional emission evolve throughout the integration.

\subsection{Application to the Didymos System}
\label{subsec:spin-orbit}

The Didymos (primary $A$) shape model has 2000 facets, while the impact-derived and relaxed Dimorphos (secondary $B$) shape models each have approximately 3000 facets (Figure~\ref{fig:model-overview}).
The impact-derived shape is reconstructed from coupled smoothed-particle hydrodynamics--discrete element method (SPH--DEM) simulations following \citet{jiao2023sph} and scaled to the JPL s547 semiaxes in Table~\ref{tab:model-parameters}.
The relaxed shape is constructed from the impact-derived mesh by smoothing boulder-scale relief and reshaping the impact depression, representing possible surface modification by post-impact landslides and granular creep \citep{agrusa2022rotation,cuk2024byorp}.
Both shape models remain fixed throughout their respective integrations and use the same dynamical masses and gravity coefficients to isolate the radiative effect of geometry, with no volume rescaling after the relaxed mesh is smoothed.
Both initially place the depression on the leading hemisphere and are evaluated with and without unresolved roughness (Section~\ref{subsec:thermal-roughness}).

\begin{table}[tbp]
\centering
\caption{Adopted Model Parameters}
\label{tab:model-parameters}
\footnotesize
\setlength{\tabcolsep}{0pt}
\newcommand{\modelparamcell}[1]{\parbox[t]{0.48\columnwidth}{\raggedright\strut#1\strut}}
\begin{tabular}{@{}l@{\hspace{0.04\columnwidth}}l@{}}
\hline\hline
\modelparamcell{Parameter} & \modelparamcell{Adopted value} \\
\hline
\multicolumn{2}{c}{\textit{Orbit and gravity}} \\
\modelparamcell{Post-transient period $P$} & \modelparamcell{$11.3674$ h} \\
\modelparamcell{Primary $Gm_A$} & \modelparamcell{$35.1175$ m$^3$ s$^{-2}$} \\
\modelparamcell{Secondary $Gm_B$} & \modelparamcell{$0.323042$ m$^3$ s$^{-2}$} \\
\modelparamcell{Initial libration angle $\phi_0$} & \modelparamcell{$0$ rad} \\
\modelparamcell{Initial libration rate $\dot\phi_0$} & \modelparamcell{$-1.22753\times10^{-5}$ rad s$^{-1}$} \\
\modelparamcell{Primary $J_{2,A}$} & \modelparamcell{$0.0899440$} \\
\modelparamcell{Primary gravity reference radius $R_{A,\mathrm{ref}}$} & \modelparamcell{$393.97$ m} \\
\modelparamcell{Secondary semiaxes $(a_B,b_B,c_B)$} & \modelparamcell{$(95.94,\ 73.80,\ 58.56)$ m} \\
\modelparamcell{Transient amplitude $A_d$} & \modelparamcell{$4.16589\times10^{-14}$ rad s$^{-2}$} \\
\modelparamcell{Decay time $\tau$} & \modelparamcell{$11.5262$ d} \\
\hline
\multicolumn{2}{c}{\textit{Thermophysical properties}} \\
\modelparamcell{Bond albedo $A_{\mathrm{B}}$} & \modelparamcell{$0.07$} \\
\modelparamcell{Emissivity $\epsilon$} & \modelparamcell{$0.90$} \\
\modelparamcell{Thermal inertia $\Gamma$ (J m$^{-2}$ K$^{-1}$ s$^{-1/2}$)} & \modelparamcell{$320[r_h/(1.1\,{\mathrm{au}})]^{-0.75}$} \\
\modelparamcell{Density $\rho$} & \modelparamcell{$2400$ kg m$^{-3}$} \\
\modelparamcell{Specific heat capacity $C_p$} & \modelparamcell{$550$ J kg$^{-1}$ K$^{-1}$} \\
\modelparamcell{Rms slope $\theta_{\mathrm{rms}}$} & \modelparamcell{$40^\circ$ (rough)} \\
\modelparamcell{Crater coverage $f_R$} & \modelparamcell{$0$ (smooth), $0.666389$ (rough)} \\
\hline
\multicolumn{2}{c}{\textit{Numerical settings}} \\
\modelparamcell{Depth steps $(\Delta z_A,\Delta z_B)$} & \modelparamcell{$(0.005,\ 0.010)$ m} \\
\modelparamcell{Depth nodes $N_{z,A},N_{z,B}$} & \modelparamcell{$20$ each} \\
\modelparamcell{Time step $\Delta t$} & \modelparamcell{$120$ s} \\
\modelparamcell{Post-impact duration $t_{\mathrm{end}}-t_0$} & \modelparamcell{$1800$ d} \\
\modelparamcell{Initial temperature $T_{\mathrm{init}}$} & \modelparamcell{$200$ K} \\
\hline
\end{tabular}
\par\vspace{3pt}
\begin{minipage}{\columnwidth}
\scriptsize\raggedright
\textit{Sources.} Orbit and gravity: JPL s547 \citep{naidu2023s547,naidu2024orbital}, with $\phi_0=0$ prescribed and the listed libration initial conditions applying to post-impact dynamical integrations.
$A_{\mathrm{B}}$, $\epsilon$, $\Gamma$, $C_p$, and $\theta_{\mathrm{rms}}$: \citet{rozitis2024preimpact}.
Density: \citet{daly2023successful,raducan2024physical}.
Thermal assumptions and the derived $f_R$ are described in Sections~\ref{subsec:binary-thermophysical}--\ref{subsec:thermal-roughness}.
\end{minipage}
\end{table}

The pre-impact comparison uses shape models reconstructed from DART imaging \citep{daly2024updated,barnouin2024geology}, with the orbit and attitudes prescribed by SPICE.
After 100 days of thermal initialization from 2020 January 1, the pre-impact analysis covers one full heliocentric revolution, from 2020 April 10 to 2022 May 20 ($\sim 770$ days).
The same thermal model follows the changing solar geometry over $r_h=1.014$--$2.275\,{\mathrm{au}}$ and solar latitudes of $-12.26^\circ$ to $+12.26^\circ$ relative to the mutual orbital plane.

For each post-impact rotational state, all four shape--surface combinations are evaluated along a common orbital and rotational history to isolate their effects on recoil \citep{quillen2022npa,sorli20253d}, particularly for tumbling, whose long-term evolution is sensitive to small perturbations \citep{gramigna2024f2bp}.
Uniform synchronous motion is specified analytically as a reference for a secondary whose rotational excitation has damped, whereas the planar-libration and tumbling histories are integrated without radiation or tides.
Uniform synchronous rotation and planar libration are both synchronous principal-axis states, with the secondary rotating about its short axis normal to the orbital plane.
Didymos's orientation follows SPICE throughout.

The uniform synchronous calculations cover 2026 November 1--2027 May 1 after 30 days of thermal initialization, using a circular orbit with a separation of 1175 m and the s547 post-transient period in Table~\ref{tab:model-parameters}.
Dimorphos rotates at the orbital rate with no libration, keeping its long axis directed toward Didymos and its depression on the leading hemisphere.

The planar dynamics retain second-degree nonspherical gravity \citep{scheeres2009stability,naidu2024orbital}, with Dimorphos's gravity coefficients and inertia derived from the homogeneous ellipsoid specified in Table~\ref{tab:model-parameters}.
The initial orbit and fitted libration rate come from JPL s547 at the DART impact ($t_0$) on 2022 September 26 \citep{naidu2023s547}.
Orbital longitude $\theta$ is measured from the initial line of centers, while $\phi$ denotes the secondary's long-axis angle from the instantaneous line of centers, with $\phi_0=0$ prescribed.
The fitted s547 exponential correction for the early post-impact transient acts only in the orbital-longitude equation, with its amplitude and decay time listed in Table~\ref{tab:model-parameters}.

The spatial integration starts from the same impact orbit and libration rate as the planar calculation, initially with Dimorphos's long axis toward Didymos and its short axis and spin along the orbit normal.
With the constraint to planar rotation removed, Dimorphos develops barrel instability, a tumbling state in which it rolls about its long axis while that axis remains approximately directed toward Didymos \citep{cuk2021barrel}.
The gravity model, empirical orbital correction, and solar and planetary perturbations follow \citet{wu2026iac}.

Separate coupled planar integrations quantify the accumulated orbital and libration-phase responses to radiative recoil.
Recoil on the components of mass $m_A$ and $m_B$ gives the relative radiative acceleration
\begin{equation}
 \boldsymbol{a}_{\mathrm{rad}}=\frac{\boldsymbol{F}_B}{m_B}-\frac{\boldsymbol{F}_A}{m_A}.
 \label{eq:relative-radiative-acceleration}
\end{equation}
Its radial and transverse components enter the mutual orbital equations, while the orbit-normal radiative torque acts on Dimorphos's spin, following the full coupled equations given by \citet{wu2026iac}.

In the coupled planar calculations, all four shape--surface cases start from the same DART state and are integrated with and without radiative recoil, omitting tides and using the numerical settings in Table~\ref{tab:model-parameters}.
To isolate radiation's subsequent effect during Hera, two continuations start from the radiative case's orbital and rotational state on 2026 November 1, retaining radiation in one and omitting it in the other.
These pairs span the adopted 181-day interval plus 15 days for centered endpoint averaging.

Separate planar integrations provide tidal comparisons with the radiative orbital and libration responses, using the torque law of \citet{cueva2024secular} with radiative recoil omitted.
Each case adopts the same $k_2/Q=4\times10^{-6}$, $10^{-5}$, or $2.5\times10^{-5}$ for both bodies, where $k_2$ is the degree-two Love number and $Q$ the tidal quality factor.
Near synchronism, the secondary torque is linearized in relative spin rate following \citet[Appendix~C]{jacobson2011fission}.
The tidal model retains the secondary's tidal spin torque and the orbital reaction torques associated with tides raised on both bodies, with paired continuations from each case's state at Hera arrival retaining or omitting tides.
Two further 1800-day integrations in planar and three-dimensional rotation retain $(k_2/Q)_A=10^{-5}$ but adopt a fixed $(Q/k_2)_B=71.6$, following \citet{burnett2025tidal}, to examine whether strong secondary dissipation damps excitation before Hera.

\section{Results and Discussion}  \label{sec:results}

The following sections quantify the effects of thermal roughness and rotational state on radiative migration and orbital phase, compare radiative and tidal responses, and discuss their implications for Hera observations and binary evolution.

\subsection{Thermal Roughness and Radiative Migration}
\label{subsec:radiative-work}

The instantaneous radiative contribution to the mutual semimajor-axis drift is
\begin{equation}
 \dot a_{\mathrm{rad}}=\frac{2a_{\mathrm{osc}}^2}{\mu}\,
 \boldsymbol{v}\boldsymbol{\cdot}\boldsymbol{a}_{\mathrm{rad}},
 \label{eq:radiative-work}
\end{equation}
where $\mu=G(m_A+m_B)$, $\boldsymbol{v}$ is relative velocity, and $a_{\mathrm{osc}}$ is the osculating two-body semimajor axis.
Primary and secondary contributions use $-\boldsymbol{F}_A/m_A$ and $\boldsymbol{F}_B/m_B$, respectively (Equation~\eqref{eq:relative-radiative-acceleration}).
Time integration gives the radiative change $\Delta a$, whose ratio to the elapsed time defines the mean rate $\langle\dot a\rangle$ over the intervals in Table~\ref{tab:radiative-budget}.
Rates are expressed in cm yr$^{-1}$ using a 365.25-day year, with negative values denoting contraction.

Using the pre-impact Didymos binary as a reference before considering impact-induced reshaping and rotational excitation, we find that roughness weakens the mean radiative contraction rate over a full heliocentric orbit from $-0.9740$ to $-0.6659\ {\mathrm{cm\,yr^{-1}}}$, a $31.6\%$ reduction in magnitude (Table~\ref{tab:radiative-budget}A).
The secondary accounts for most of the contraction and more than $99\%$ of the roughness-induced change, with its contribution reduced in magnitude by $32.2\%$, compared with $7.4\%$ for the primary.
The primary's total radiative contribution amounts to $37$--$40\%$ of the $\simeq-0.05\ {\mathrm{cm\,yr^{-1}}}$ eclipse-driven BYP estimated by \citet{zhou2024binary}.
Their estimate assumes spherical components and coplanar mutual and heliocentric orbits, whereas our calculation includes resolved shape and seasonal illumination.
Their Appendix~A also predicts weaker eclipse-driven recoil when these planes are inclined.

For the adopted pre-impact shapes, the total radiative contraction rate is $8$--$12$ times the magnitude of the observed net drift, $-0.08\pm0.02\ {\mathrm{cm\,yr^{-1}}}$ over 2003--2022 \citep{scheirich2024dimorphos,zhou2024binary}.
Orbital expansion driven by tides raised on the rapidly rotating primary can offset most of this contraction, leaving a small net drift \citep{jacobson2011equilibria,cueva2024secular}.

Roughness modifies the directionality of thermal emission from differently heated crater walls \citep{rozitis2011directional}, so its effect on migration depends on their combined recoil along the relative velocity rather than on total infrared power.
To examine how the reduction found for Didymos depends on shape and illumination, we use an idealized binary with a spherical primary and a synchronously rotating secondary formed by flattening a cap of angular radius $30^\circ$ at the leading equatorial point of a sphere and rescaling the body to restore its volume.
The components have volume-equivalent radii of 500 and 100 m, a separation of 1500 m, and a density of $2000\ {\mathrm{kg\,m^{-3}}}$.
Calculations use $\Gamma=200\ {\mathrm{J\,m^{-2}\,K^{-1}\,s^{-1/2}}}$, $A_{\mathrm{B}}=0.1$, $\epsilon=0.9$, and irradiance $1361\ {\mathrm{W\,m^{-2}}}$, with a fixed inertial solar direction and no eclipses or interbody irradiation to isolate shape-dependent recoil.

At zero solar latitude, roughness weakens both inward and outward migration: increasing crater coverage from $f_R=0$ to 1 changes the secondary's contribution from $-1.505$ to $-0.776\ {\mathrm{cm\,yr^{-1}}}$ for the leading cap.
The corresponding change is $+1.066\to+0.550\ {\mathrm{cm\,yr^{-1}}}$ for a trailing-cap configuration obtained by rotating the cap $135^\circ$ about the positive orbit normal through the primary-facing side.
At higher solar latitudes, however, the same increase in roughness strengthens inward migration for the original leading-cap shape at $42.5^\circ$ ($-0.120\to-0.222\ {\mathrm{cm\,yr^{-1}}}$) but reverses outward migration at $45^\circ$ ($+0.070\to-0.254\ {\mathrm{cm\,yr^{-1}}}$).

To isolate roughness effects on eclipse-driven recoil, we adopt spherical components while retaining the radii, separation, density, and thermophysical parameters above.
Either the primary or the secondary spins at five times the mutual orbital mean motion while its companion remains synchronous.
Relative to smooth surfaces, full crater coverage ($f_R=1$) at $\Gamma=200$ strengthens contraction from the rapidly rotating secondary (BYS) by $10.1\%$ but weakens that from the rapidly rotating primary (BYP) by $3.2\%$.
These contributions are isolated by comparing otherwise identical calculations with and without eclipses at zero solar latitude, with interbody irradiation omitted.
These cases show that thermal roughness can weaken, strengthen, or reverse radiative migration by changing the balance of recoil contributions to orbital work, with the outcome depending on shape, illumination, and rotation.

\subsection{Post-impact Spin--Orbit Evolution}
\label{subsec:phase-accumulation}

The post-impact calculations compare radiative migration along prescribed orbital and rotational histories and quantify the accumulated orbital-phase contribution through separate coupled planar integrations.
Over the common six-month Hera interval used for the migration-rate comparisons, roughness reduces the mean radiative contraction rate by $28$--$39\%$ relative to smooth surfaces in synchronous principal-axis rotation (Table~\ref{tab:radiative-budget}B).

For uniform synchronous rotation, the secondary's dimensionless BYORP coefficient is
\begin{equation}
 B=\frac{\langle\boldsymbol{F}_B\!\boldsymbol{\cdot}\!\boldsymbol{e}_\theta\rangle}
 {R_V^2\langle P_\odot\rangle},
 \label{eq:byorp-coefficient}
\end{equation}
where $\boldsymbol{e}_\theta$ is the prograde transverse unit vector, $P_\odot$ is solar radiation pressure at the system's heliocentric distance, and angle brackets denote averages over one mutual orbit \citep{mcmahon2010secular}.
Negative $B$ denotes inward BYORP migration, whose characteristic timescale scales as $|B|^{-1}$ for fixed system parameters and solar forcing.
These orbit-by-orbit coefficients are averaged over the Hera interval.
Both shapes use $R_V=74.89\ {\mathrm{m}}$, the volume-equivalent radius of the DART-derived pre-impact Dimorphos shape, following the area normalization of \citet{cueva2024secular}.
Here $B$ characterizes the secondary's transverse recoil, including thermal response, eclipses, and mutual heating, whereas $\langle\dot a\rangle$ also includes primary recoil and radial work.
Roughness reduces $|B|$ by $33.7\%$ for the impact-derived shape and $26.7\%$ for the relaxed shape (Table~\ref{tab:radiative-budget}B).
With roughness included, the two shapes yield $|B|=0.01347$ and $0.02964$, of the same order as the post-impact range $|B|\simeq0.036$--$0.045$ implied by the nominal global-reshaping case of \citet{cueva2024secular}.

Over the same Hera interval, recoil evaluated along the prescribed planar-libration trajectory with surface roughness included gives mean radiative contributions to the migration rate of $-1.917$ and $-4.649\ {\mathrm{cm\,yr^{-1}}}$ for the impact-derived and relaxed shapes, respectively.
The relaxed shape produces approximately $2.4$ times the radiative contraction of the impact-derived shape along the same planar trajectory, showing that the adopted reshaping strengthens inward migration even though unresolved thermal roughness weakens it for either fixed shape.
These planar radiative contributions are comparable in magnitude to the combined tidal--BYORP rates of $-1.8$ to $-4.6\ {\mathrm{cm\,yr^{-1}}}$ found by \citet{cueva2024secular} for their nominal post-impact shapes undergoing libration.
The modeled barrel-instability state gives much weaker net radiative migration: with surface roughness included, its contraction rates are only $13$--$19\%$ of the corresponding planar-libration values, while roughness reduces contraction by $16$--$18\%$ relative to smooth surfaces.

\begin{figure*}[t]
\centering
\includegraphics[trim=16.80bp 0 26.52bp 0,clip,width=0.80\textwidth]{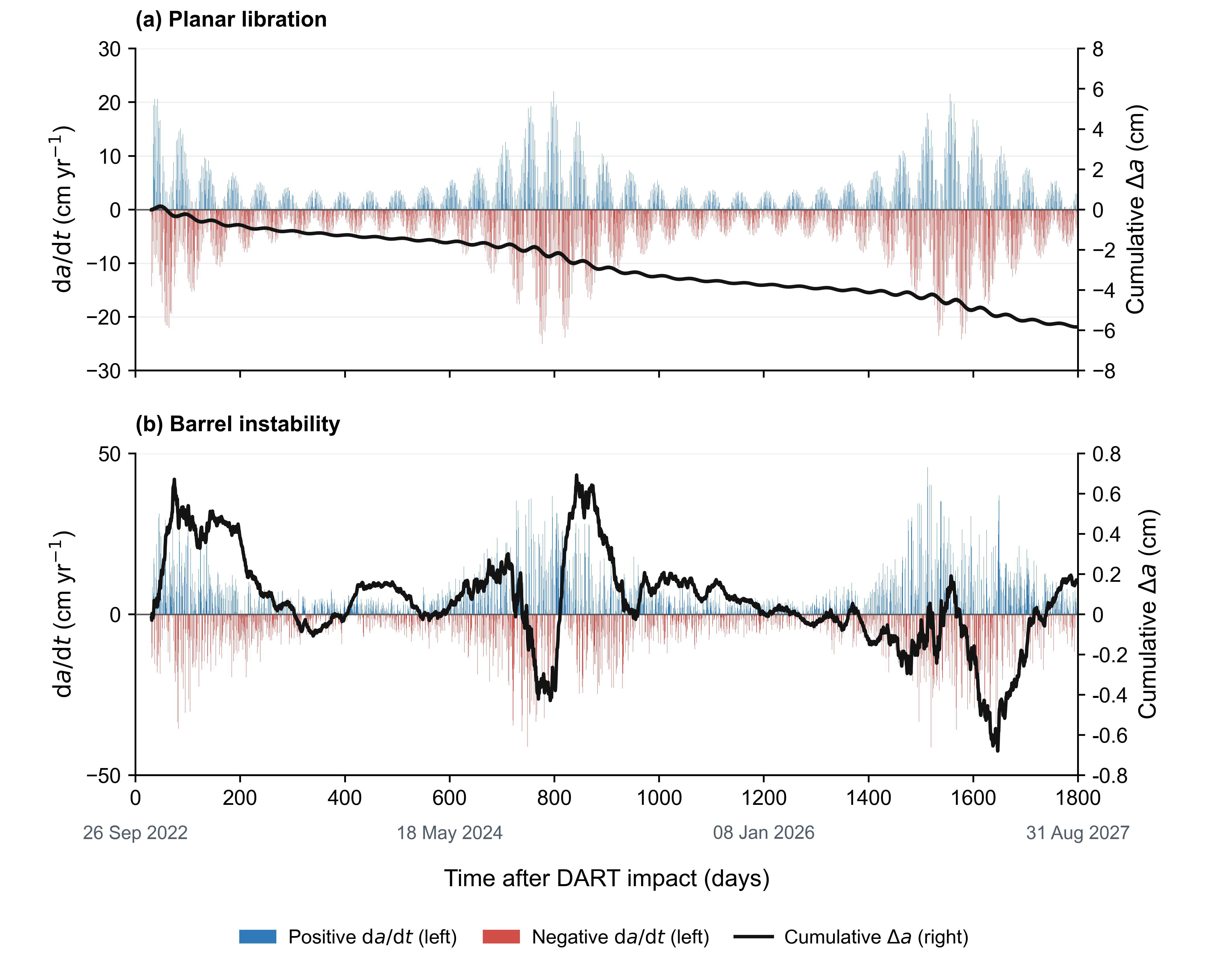}
\caption{Radiative migration for the impact-derived shape with a rough surface over days 30--1800 after DART, following thermal initialization: (a) planar libration and (b) barrel instability.
Bars show the radiative contribution to $\mathrm{d}a/\mathrm{d}t$ averaged over one complete mutual orbit, while black curves show cumulative $\Delta a$ from radiative work, with different vertical scales in the two panels.
Recoil is evaluated along trajectories integrated without radiation or tides.}
\label{fig:recoil-accumulation}
\end{figure*}

Over days 30--1800 after DART, integrating radiative work along the prescribed planar-libration trajectory gives $\Delta a=-5.837\ {\mathrm{cm}}$ for the impact-derived shape with surface roughness included, despite frequent sign changes in the orbit-averaged drift (Figure~\ref{fig:recoil-accumulation}).
Along the barrel-instability trajectory, the net radiative contribution over the same span is much smaller at $\Delta a=+0.170\ {\mathrm{cm}}$, even though the mean drift is inward during the adopted Hera interval (Table~\ref{tab:radiative-budget}B).

\begin{figure*}[t]
\centering
\includegraphics[width=0.95\textwidth]{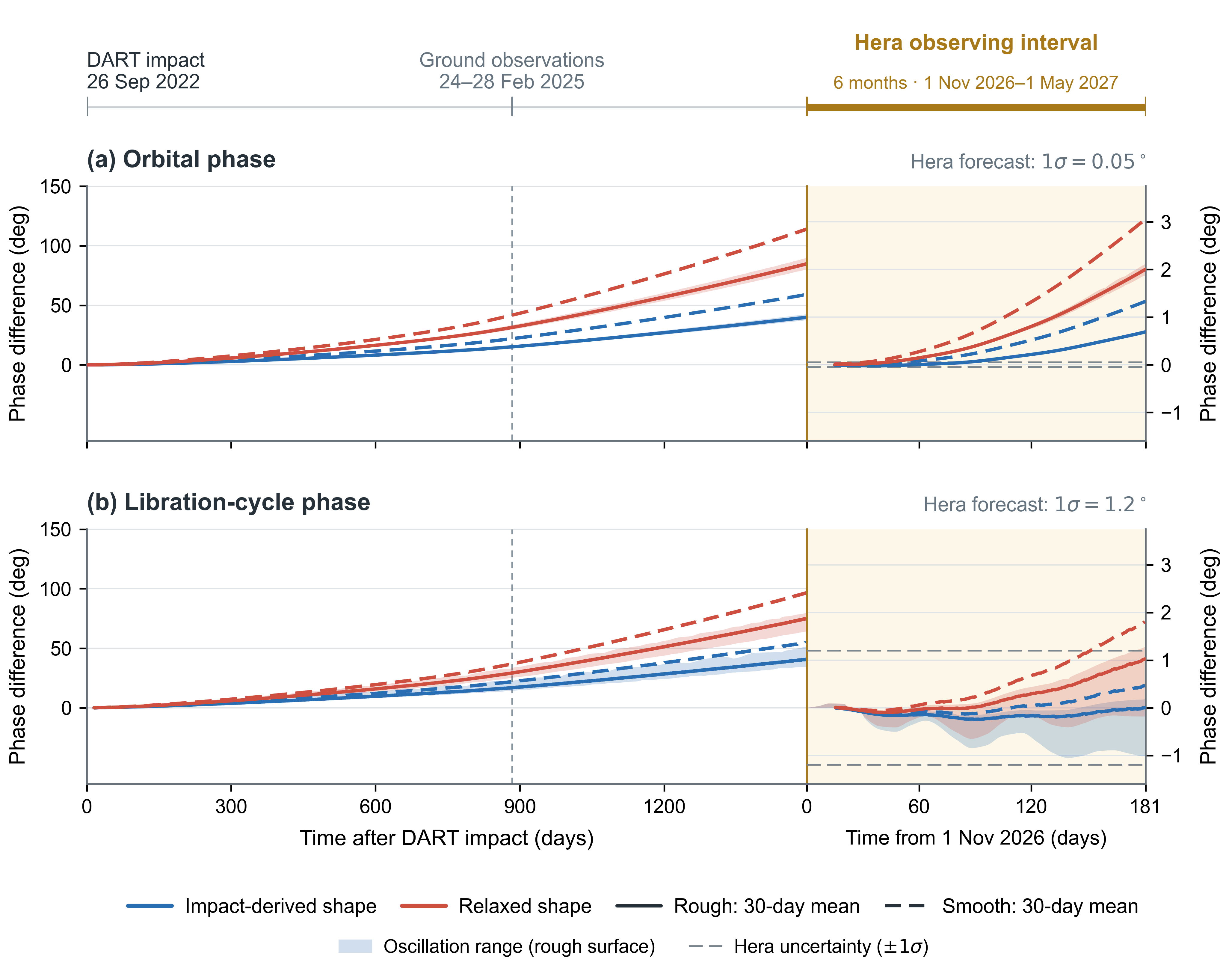}
\caption{Radiative contributions to (a) orbital longitude and (b) libration-cycle phase ($360^\circ$ per oscillation), defined as radiation on minus radiation off in tide-free coupled planar integrations.
The left segments start from the common DART-impact state, while the right segments cover 2026 November 1--2027 May 1, with each pair sharing the state reached with recoil at Hera arrival.
Solid and dashed curves show centered 30-day means for rough and smooth surfaces, while shading bounds the oscillations of the rough-surface phase differences.
The timeline marks February 2025 observations \citep{knight2025ground,naidu2026updated}.
Gray horizontal lines show published $\pm1\sigma$ formal uncertainties for comparison: orbital longitude, converted from along-track position (a), and fitted sinusoidal libration phase (b) \citep{gramigna2024hera}.}
\label{fig:coupled-phase}
\end{figure*}

In separate coupled planar-libration integrations, radiative recoil feeds back on the mutual orbit and Dimorphos's rotation, producing the phase responses in Figure~\ref{fig:coupled-phase} and Table~\ref{tab:radiative-budget}C.
Orbital and libration-cycle phase differences are defined as radiation on minus radiation off and averaged over complete, centered 30-day windows, with $\Delta\theta_{30}$ denoting the mean orbital-phase difference.
Libration-cycle phase increases linearly by $360^\circ$ between successive positive-to-negative zero crossings of $\phi$ and accumulates across cycles, measuring oscillation timing rather than the physical libration angle.
For the Hera continuations, orbital-phase means are reported at day 181, while cycle-based libration means are reported at day 180, the last integer day with complete centered windows for every case.

In these coupled calculations, centimeter-scale semimajor-axis contraction slightly shortens the orbital period, allowing a small mean-motion difference to accumulate into substantial phase advances over successive revolutions.
Over the roughly four years from DART to Hera's arrival, radiative recoil contributes mean orbital-phase advances of approximately $40^\circ$ and $85^\circ$ for the impact-derived and relaxed shapes with rough surfaces, respectively, $32.4\%$ and $25.6\%$ below the corresponding smooth-surface values.
In paired continuations sharing each case's state at Hera arrival, radiative recoil contributes mean orbital-phase advances of approximately $0.7^\circ$ and $2.0^\circ$ over the six-month investigation, compared with $1.3^\circ$ and $3.1^\circ$ for smooth surfaces (Figure~\ref{fig:coupled-phase}a and Table~\ref{tab:radiative-budget}C).

Full two-body covariance analyses support joint estimation of Dimorphos's orbit and rotation over short observation arcs even in non-principal-axis states \citep{gramigna2024f2bp}, with CubeSat imaging providing additional constraints \citep{gramigna2025cubesat}.
In the coupled Hera continuations, roughness reduces the radiative orbital-phase advance by $0.64^\circ$ and $1.05^\circ$ for the impact-derived and relaxed shapes, respectively.
Both the radiative phase advances and these differences lie well above the $0.05^\circ$ formal orbital-longitude uncertainty converted from the upper $100\ {\mathrm{cm}}$ along-track uncertainty reported by \citet{gramigna2024hera}.
By contrast, the day-180 mean radiative contributions to libration-cycle phase for the impact-derived and relaxed shapes with surface roughness included, $-0.0006^\circ$ and $1.016^\circ$, are smaller in magnitude than the published $1.20^\circ$ formal uncertainty for a fitted sinusoidal libration phase, used here for an approximate comparison with the cycle-based phase.

Ground-based observations in 2025 recorded four mutual events approximately 70--90 minutes earlier than predicted by the published 2024 orbit solutions, a $1.5$--$1.6\sigma$ difference over 29 months, although a slightly shorter constant period also fits these timings \citep{knight2025ground,naidu2026updated}.
At the 2025 epoch, the coupled planar calculations with surface roughness included yield mean radiative phase contributions equivalent to timing advances of approximately 29 and 59 minutes for the impact-derived and relaxed shapes, respectively (Table~\ref{tab:radiative-budget}C).
The comparable magnitudes suggest that radiative recoil may have contributed appreciably to orbital contraction, alongside secondary tidal damping or other dissipation of Dimorphos's residual excitation, counteracting the outward contribution from primary tides \citep{meyer2023energy,richardson2024dynamical,cuk2024byorp}.
Hera's measurements of Dimorphos's orbital phase at arrival, compared with predictions updated using the 2025 observations, could offer further clues to the system's post-impact evolution \citep{naidu2026updated}.

\begin{table*}[t]
\caption{Orbital Migration and Phase Response}
\label{tab:radiative-budget}
\footnotesize
\setlength{\tabcolsep}{0pt}
\renewcommand{\arraystretch}{1.13}
\par\noindent\makebox[\textwidth][c]{%
\begin{minipage}{0.94\textwidth}
\centering
\begin{tabular*}{\linewidth}{@{\extracolsep{\fill}}lccc@{}}

\multicolumn{4}{l}{\textbf{A. Pre-impact radiative migration along prescribed motion: $\langle\dot a\rangle$ (cm yr$^{-1}$)}} \\
\hline
\makebox[0.32\linewidth][l]{Surface} & \makebox[0.2266\linewidth][c]{Primary (Didymos)} & \makebox[0.2266\linewidth][c]{Secondary (Dimorphos)} & \makebox[0.2266\linewidth][c]{Total} \\
\hline
Smooth & $-0.0202$ & $-0.9538$ & $-0.9740$ \\
Rough  & $-0.0187$ & $-0.6471$ & $-0.6659$ \\
\hline
\end{tabular*}
\par\vspace{5pt}
\begin{tabular*}{\linewidth}{@{\extracolsep{\fill}}llcccc@{}}
\multicolumn{6}{l}{\textbf{B. Post-impact migration contributions}} \\
\multicolumn{6}{l}{2026 Nov 1--2027 May 1 (rates in cm yr$^{-1}$)} \\
\multicolumn{6}{l}{\textit{Radiative contributions}} \\
\hline
\makebox[0.20\linewidth][l]{Shape model} & \makebox[0.12\linewidth][l]{Surface} & \multicolumn{2}{c}{Uniform synchronous rotation} & Planar libration & Barrel instability \\
 & & \makebox[0.17\linewidth][c]{$B$} & \makebox[0.17\linewidth][c]{$\langle\dot a\rangle$} & \makebox[0.17\linewidth][c]{$\langle\dot a\rangle$} & \makebox[0.17\linewidth][c]{$\langle\dot a\rangle$} \\
\hline
Impact-derived & Smooth & $-0.02034$ & $-4.376$ & $-3.151$ & $-0.446$ \\
Impact-derived & Rough  & $-0.01347$ & $-2.861$ & $-1.917$ & $-0.365$ \\
Relaxed     & Smooth & $-0.04046$ & $-8.780$ & $-6.599$ & $-0.739$ \\
Relaxed     & Rough  & $-0.02964$ & $-6.342$ & $-4.649$ & $-0.620$ \\
\hline
\end{tabular*}\par
\begin{tabular*}{\linewidth}{@{\extracolsep{\fill}}lccc@{}}
\multicolumn{4}{l}{\textit{Primary tidal orbital work along the planar trajectory}} \\
\makebox[0.32\linewidth][l]{Primary tidal strength $(k_2/Q)_A$} & \makebox[0.2266\linewidth][c]{$4\times10^{-6}$} & \makebox[0.2266\linewidth][c]{$10^{-5}$} & \makebox[0.2266\linewidth][c]{$2.5\times10^{-5}$} \\
Primary tidal $\langle\dot a\rangle$ (cm yr$^{-1}$) & $+0.182$ & $+0.454$ & $+1.135$ \\
\hline
\end{tabular*}
\par\vspace{5pt}
\begin{tabular*}{\linewidth}{@{\extracolsep{\fill}}llccc@{}}
\multicolumn{5}{l}{\textbf{C. Radiative orbital-phase advance from coupled planar integrations: $\Delta\theta_{30}$ (deg)}} \\
\multicolumn{5}{l}{Centered 30-day means evaluated at the indicated epochs} \\
\hline
 & & \multicolumn{2}{c}{Since DART} & Since Hera arrival \\
\makebox[0.20\linewidth][l]{Shape model} & \makebox[0.12\linewidth][l]{Surface} & \makebox[0.2266\linewidth][c]{2025 Feb 26} & \makebox[0.2266\linewidth][c]{2026 Nov 1} & \makebox[0.2266\linewidth][c]{2027 May 1} \\
\hline
Impact-derived & Smooth & $+21.943$ & $+59.029$ & $+1.333$ \\
Impact-derived & Rough  & $+15.093$ & $+39.900$ & $+0.692$ \\
Relaxed     & Smooth & $+41.686$ & $+113.974$ & $+3.061$ \\
Relaxed     & Rough  & $+31.372$ & $+84.802$ & $+2.006$ \\
\hline
\end{tabular*}\par
\vspace{4pt}
\scriptsize\raggedright
\textit{Notes.} Rates use a 365.25-day year, with negative values denoting contraction.
A averages one heliocentric revolution, 2020 Apr 10--2022 May 20 ($\sim 770$ days).
Radiative rates in B are averaged over complete mutual orbits within the stated Hera interval, using a common trajectory for all shape--surface cases within each rotational state.
The secondary's BYORP coefficient $B$ uses the common normalization in Equation~\eqref{eq:byorp-coefficient}, whereas radiative $\langle\dot a\rangle$ includes both bodies.
Primary tidal rates are calculated from orbital work along the same planar trajectory and averaged over all 181 days.
The coupled tidal integrations in Section~\ref{subsec:dissipation-hera} include tides on both bodies.
In C, phase contributions compare integrations with and without radiation, sharing the DART-impact state in the ``Since DART'' columns and the radiative case's arrival state in ``Since Hera arrival.''

\end{minipage}%
}\par
\end{table*}

\subsection{Dissipation and Dynamical Relaxation}
\label{subsec:dissipation-hera}

Tides raised on Didymos, whose spin rate exceeds the mutual mean motion, transfer spin angular momentum to the orbit, providing the outward contribution that can balance inward BYORP \citep{jacobson2011equilibria}.
Applying Equation~\eqref{eq:radiative-work} to the primary-tide acceleration along the common planar trajectory gives a six-month mean contribution of $+0.454\ {\mathrm{cm\,yr^{-1}}}$ at $(k_2/Q)_A=10^{-5}$, scaling linearly to $+0.182$--$+1.135\ {\mathrm{cm\,yr^{-1}}}$ over the three strengths in Table~\ref{tab:radiative-budget}.
Roughness weakens the planar radiative contraction by $1.234$ and $1.950\ {\mathrm{cm\,yr^{-1}}}$ for the impact-derived and relaxed shapes, respectively, about $2.7$ and $4.3$ times the primary-tide contribution at $(k_2/Q)_A=10^{-5}$.

In integrations with tides on both bodies, secondary tides damp libration and modify the orbit through spin--orbit coupling \citep{meyer2023energy}, partially offsetting the phase delay from primary tides.
Relative to a tide-free continuation from the Hera-arrival state of the $k_2/Q=10^{-5}$ tidal case, the day-181 mean orbital-phase shifts are $-0.202^\circ$ for primary tides alone and $+0.173^\circ$ for secondary tides alone, compared with $-0.029^\circ$ for both.
Across the three strengths, the paired Hera integrations with tides on both bodies yield 30-day mean orbital-phase shifts of $-0.011^\circ$ to $-0.087^\circ$ at day 181.

Although their orbital-phase shifts are small, the same tidal integrations give 30-day mean libration-cycle phase advances of $3.77^\circ$--$20.68^\circ$ at day 180 relative to the paired tide-free continuations.
The corresponding reductions in libration amplitude, measured as half the peak-to-peak angle per cycle, are $0.0235^\circ$--$0.1291^\circ$, comparable to or larger than Hera's projected $0.0182^\circ$ formal uncertainty in fitted sinusoidal libration amplitude \citep{gramigna2024hera}.
With surface roughness included, the radiative amplitude changes remain below $0.0017^\circ$ in absolute value over all complete 30-day windows and are unlikely to be resolved at this projected precision.
In these planar calculations, orbital phase responds strongly to radiation whereas libration amplitude is more sensitive to tidal damping.

Efficient dissipation, potentially enhanced by the episodic landsliding proposed by \citet{cuk2024byorp}, could bring Dimorphos close to uniform synchronous rotation before Hera arrives and thereby change the radiative contribution to orbital evolution.
By explicitly resolving contact and friction between constituent particles, \citet{burnett2025tidal} simulated efficient tidal spin-down in a rubble-pile secondary and obtained a mean effective $Q/k_2=71.6$.
With this mean held fixed as $(Q/k_2)_B$ and $(k_2/Q)_A=10^{-5}$ retained, rotational excitation damps out within months after DART in both the planar and three-dimensional integrations.
In the idealized circular-orbit calculations with uniform synchronous rotation and surface roughness included, the combined recoil of both bodies contributes $-2.861$ and $-6.342\ {\mathrm{cm\,yr^{-1}}}$ to the mean migration rates for the impact-derived and relaxed shapes, respectively, exceeding the corresponding planar-libration contributions in magnitude by $49\%$ and $36\%$ (Table~\ref{tab:radiative-budget}B).

After excitation decays, the final orientation still affects recoil because second-degree gravity admits two equivalent long-axis alignments $180^\circ$ apart, exposing different surfaces at the same orbital phase.
A trailing-side depression could reverse the shape-dependent BYORP contribution relative to the leading-side depression in Table~\ref{tab:radiative-budget}, although the total rates need not be equal and opposite because eclipses, thermal inertia, mutual heating, and primary recoil also affect the drift.
Hera's shape and orientation measurements are expected to resolve this ambiguity.

\subsection{Implications for Hera Observations}
\label{subsec:hera-observations}

\citet{rozitis2024preimpact} inferred an rms slope of $40^\circ\pm3^\circ$ and $\Gamma=320\pm70\ {\mathrm{J\,m^{-2}\,K^{-1}\,s^{-1/2}}}$ from the unresolved pre-impact system, whose thermal flux is dominated by Didymos.
Changes in Dimorphos's thermal properties following impact excavation and redistribution \citep{jiao2023sph,raducan2024physical,daly2024updated} may therefore be difficult to identify in unresolved infrared observations.
By resolving the two bodies at different rotational phases and viewing angles, Hera's Thermal InfraRed Imager (TIRI) will provide the observations needed to estimate each body's thermal inertia and roughness for BYORP calculations \citep[Section~5.5]{okada2025tiri}.

Combining these thermal constraints with Hera's surface mapping and measurements of Dimorphos's post-impact orbit and rotation \citep{naidu2024orbital,richardson2024dynamical,tortora2025radio} would allow radiative recoil to be calculated for the observed system.
With this radiative contribution accounted for, joint orbital and rotational observations could constrain post-impact dissipation by quantifying the angular momentum transferred from Didymos's spin to the orbit by tides and the mechanical energy dissipated as Dimorphos's impact-induced excitation decays (Section~\ref{subsec:dissipation-hera}).
Hera's measurements of mass distribution and internal structure would help interpret these transfers and losses in terms of how readily the rubble-pile bodies deform and dissipate energy \citep{meyer2023energy,michel2022esa}, properties that also influence their response to kinetic impacts in planetary defense \citep{raducan2024physical}.
Hera's combined constraints on shape and mechanical properties could therefore inform kinetic-impact strategies, including off-center targeting to enhance deflection through ejecta recoil \citep{lee2025incremental,lee2026trajectories}, and help extend the lessons from DART to other asteroid targets.

\subsection{Thermal Roughness in Binary Evolution}
\label{subsec:binary-evolution}

Thermophysical models refined with Hera's spatially resolved thermal and shape measurements could improve estimates of unresolved roughness from disk-integrated infrared observations of other binaries \citep{okada2025tiri,rozitis2024preimpact}.
Incorporating these roughness constraints into BYORP estimates for synchronous secondaries would allow the measured orbital drift to constrain the tidal contribution.

The nearly stationary orbit of 1996~FG$_3$ has been interpreted as a balance between tides and BYORP, allowing its tidal properties to be constrained from an estimate of the radiative torque \citep{scheirich2015fg3}.
In this equilibrium, the required primary tidal strength scales as $(k_2/Q)_A\propto|B|$ for fixed system parameters and solar forcing \citep{jacobson2011equilibria}.
If radiation and primary tides account for Didymos's pre-impact migration, including thermal roughness lowers the primary tidal strength inferred from the same measured drift by about $34\%$ at fixed shape and tidal law.
Subtracting the radiative rates in Table~\ref{tab:radiative-budget}A from the measured drift of approximately $-0.08\ {\mathrm{cm\,yr^{-1}}}$ over 2003--2022 \citep{scheirich2024dimorphos,zhou2024binary} gives required outward primary-tide contributions of $+0.894\ {\mathrm{cm\,yr^{-1}}}$ for a smooth surface and $+0.586\ {\mathrm{cm\,yr^{-1}}}$ for a rough surface.
Hera is expected to refine these shape-dependent estimates by mapping the hemisphere not resolved by DART and improving the pre-impact shape reconstruction \citep{daly2024updated,cueva2024secular}.

The binary-survival simulations of \citet{cueva2026survival} show that tidal--BYORP equilibrium can allow a satellite formed in the main belt to survive Didymos's migration into near-Earth space.
For each post-impact shape, thermal roughness reduces radiative contraction by $28$--$39\%$ in synchronous principal-axis rotation (Table~\ref{tab:radiative-budget}B), lengthening the characteristic migration timescale.

Over a binary's lifetime, the cumulative radiative contribution depends on both the migration rates in different rotational states and the time spent in each state.
Asynchronous secondaries are observed \citep{pravec2016secondary}, including that of 1991~VH, whose rotation was likely excited by a planetary encounter \citep{meyer2024encounter}.
Even within the 1:1 resonance, non-principal-axis rotation can develop \citep{tan2023attitude}, weaken or reverse BYORP \citep{quillen2022npa}, and evade detection in lightcurves \citep{cuk2021barrel}.
Migration can nevertheless continue through the binary Yarkovsky effect on a rapidly rotating primary or an asynchronously rotating secondary \citep{zhou2024yarkovsky,zhou2024binary}.
If secondary dissipation damps excitation well before the next disturbance, the ensuing synchronous evolution depends on the strength and direction of radiative recoil and its balance with primary tides.
For such rapidly rotating primaries, weak recoil or cancellation between inward BYORP and outward primary tides can prolong binary lifetimes, whereas outward BYORP reinforces tidal expansion \citep{jacobson2011equilibria,cueva2026survival}.
Expansion can also amplify libration until synchronism is lost, producing wide asynchronous binaries \citep{jacobson2014wide}.

For binaries formed by rotational fission \citep{jacobson2011fission}, thermal roughness may also influence the outcomes of subsequent mass-shedding episodes through its effect on radiative migration.
In the multi-generation model of \citet{dai2026multigeneration}, the extent of outward migration before the next mass-shedding episode helps determine whether an existing satellite accretes new debris, encounters another satellite, or remains separated from it.
Relating the observed abundance and diversity of binaries to their formation and survival therefore requires roughness-dependent migration timescales, considered alongside rotational-relaxation times and the interval between mass-shedding episodes.

\section{Conclusions}  \label{sec:conclusions}

Thermal roughness significantly modifies the radiative contribution to binary orbital migration.
For the adopted post-impact Dimorphos shapes, roughness reduces the mean radiative contraction rate over the Hera interval by $28$--$39\%$ in synchronous principal-axis rotation and $16$--$18\%$ in the modeled tumbling state, relative to smooth surfaces.
Idealized configurations show that roughness can also strengthen or reverse migration as shape, illumination, and rotation vary.

The rotational state governs how radiative recoil accumulates, with much weaker net migration along the modeled tumbling trajectory than in planar libration.
With roughness included, centimeter-scale contraction in the coupled planar calculations shortens the orbital period, yielding mean radiative contributions to orbital-phase advance of $40^\circ$--$85^\circ$ from DART to Hera and a potentially detectable $0.7^\circ$--$2.0^\circ$ during Hera's six-month investigation.
The modeled tides affect libration amplitude much more strongly than radiation, making orbital and rotational observations complementary probes of recoil and dissipation.

The model treats unresolved thermal roughness and radiative exchange during rotational excitation, allowing recoil estimates to be updated with Hera's measurements of shape, thermal properties, and rotational state.
Combined with orbital and rotational observations, these estimates could constrain post-impact dissipation and rubble-pile mechanical properties relevant to planetary defense.
Beyond Didymos, roughness-dependent BYORP estimates can improve inferences of tidal response from orbital observations and clarify how tides and radiation govern the evolution and lifetimes of binary populations.

\begin{acknowledgments}
This work is supported by the National Natural Science Foundation of China under Grant Nos.
62227901, 12572404, 125B1015, and 12373066.
B.C.
is also supported by the Beijing Nova Program (No. 20250484831) and the Tsinghua Dushi Funds (No. 20261080037).
\end{acknowledgments}


\software{COBIRD v1.0.0 \citep{wu2026cobird}}

\bibliography{sample701}{}
\bibliographystyle{aasjournalv7}

\end{document}